\documentclass[12pt]{article}
\usepackage{graphicx, epsfig}
\usepackage{amsmath}
\usepackage{latexsym}
\usepackage{amstext}
\usepackage{ulem}
\usepackage{amssymb}
\usepackage{array}
\usepackage{multirow}
\usepackage{tikz}
\usetikzlibrary{arrows}
\newcommand{\mb}[1]{ \mbox{\boldmath$#1$} }

\newcommand{\beq}{\begin{eqnarray}}
\newcommand{\eeq}{\end{eqnarray}}
\newcommand{\beqq}{\begin{eqnarray*}}
\newcommand{\eeqq}{\end{eqnarray*}}

\newcommand{\X}{\mbox{\boldmath$X$}}

\newcommand{\w}{\mbox{\boldmath$w$}}

\font\bb=msbm10 at 12pt

\def\eE{\hbox{\bb E}}

\begin{document}

\pagestyle{plain}
\begin{center}
{{\large \textbf{{Analysis of single particle trajectories: when things go wrong}}}\\[5mm]
D. Holcman$^{*}$ \footnote{Ecole Normale Sup\'erieure, 46 rue d'Ulm 75005 Paris, France.}, N. Hoze$^{*}$$^1$, and Z. Schuss$^{*}$   \footnote{Department of Applied Mathematics, Tel-Aviv University, Tel-Aviv 69968, Israel}}
\\\small{$^{*}$corresponding authors: david.holcman@ens.fr, hoze@biologie.ens.fr and schuss@post.tau.ac.il }

\end{center}
\date{}
To recover the long-time behavior and the statistics of molecular trajectories from the large number
(tens of thousands) of their short fragments, obtained by super-resolution methods at the single
molecule level  \cite{Manley,Giannone}, data analysis based on a stochastic model of their
non-equilibrium motion is required. Recently, we characterized the local biophysical properties
underlying receptor motion based on coarse-grained long-range interactions, corresponding to
attracting potential wells of large sizes \cite{HozePNAS}, as predicted theoretically in
\cite{holcman2004,taflia} (see also \cite{Saxton-95} for corrals). The purpose of this letter is to
point out what was done and not done in \cite{Masson} on this subject.\\[1mm]
\textbf{Parameter estimation}\\[1mm]
The new paper of Masson et al. \cite{Masson} presents a comparable study {of} glycine
receptors at inhibitory synapses. {It is our aim here to clarify some issues that arise
from the analysis and simulations presented there}. In particular{,}  \cite{Masson}
asserts that{``}\textit{the method used in Hoze et al. (9), also based on a combination
of high-density single-molecule imaging and statistical inference, evaluates the diffusion and drift
by computing the maximal likelihood estimation in a mesh square as described in T\"urkcan et al.
(7)}." This statement {may be misleading, because the analysis of \cite{HozePNAS} does
not concern ``}maximal likelihood estimation." The analysis of \cite{HozePNAS} is based on the
assumption that the molecular motion is a diffusion process, whose local drift and diffusion
coefficients are estimated from local conditional moments of the available many fragments of the
trajectories. The estimates in \cite{HozePNAS} are the standard non-parametric empirical estimates
of the drift and diffusion coefficients known for decades \cite{Karlin,Schuss1,Schuss2,Schuss3}.
Thus the analysis presented in  T\"urkcan et al \cite{Turkcan}, which was actually published after \cite{HozePNAS} has been sent for publication, {does not directly concern the issue at hand.}

The physical model of a receptor motion on a homogenous surface is the overdamped limit of the
Saffman-Delbr\"uck-Langevin model \cite{Saffman-Delbruck}, \cite{Saffman}. Specifically, the diffusion of a receptor embedded in a membrane surface is generated by a diffusion coefficient $D$ and a field of force $\mb{F}(\X)$,
\beq \label{stochlocal0}
\dot{\X}=\frac{F(\X)}{\gamma}+ \sqrt{2D}\, \dot{\w},
\eeq
where $\dot\w$ is a vector of independent standard $\delta$-correlated Gaussian white noises and
$\gamma$ is the dynamical viscosity \cite{Schuss3}. {However, the trajectory fragments
are not collected on the microscopic time-scale of  \eqref{stochlocal0}, but rather on a coarser
time-scale of the recording apparatus that coarse-grains short events that are due to crowding
organization of a variety of obstacles. Thus \eqref{stochlocal0} is coarse-grained on a coarser
spatiotemporal scale into an effective stochastic equation as (see \cite{Hoze2012})
\begin{align}\label{stochlocal01}
\dot{\X}=\mb{b}(\X) +\sqrt{2}\mb{B}_e(\X)\, \dot{\w},
\end{align}
with the empirical drift field $\mb{b}(\X)$ and diffusion matrix $\mb{B}_e(\X)$, where the effective
diffusion tensor $\mb{D}_e(\X)=\frac12\mb{B}_e(\X)\mb{B}_e^T(\X)$ can be expressed in terms of the
more microscopic diffusion coefficient $D$  and in terms of the density and geometry of obstacles.
The observed effective diffusion tensor needs not be isotropic and can be state dependent, whereas
the friction coefficient $\gamma$ remains constant. Obviously, the impenetrable obstacles that slow
the effective diffusion down affect neither the microscopic physical properties of the diffusing
particle nor those of the membrane. Note that the effective field $\mb{b}(\X)$ may have no potential
(see \cite{Schuss2}).}

Theoretically, the coefficients of \eqref{stochlocal0} can be statistically estimated at each point
$\X=(X^1,X^2)^T$ of the membrane from an infinitely large sample of its trajectories in the
neighborhood of the point $\X$ at time $t$. Specifically, setting $\Delta\X(t)=\X(t+\Delta t)-\X(t)$, the
infinitesimal coefficients are given by the limits of the expectations \cite{Schuss2}, \cite{Schuss3}
\beq
 \label{eq:force}
\mb{b}(\X)=\lim_{\Delta t \rightarrow 0} \frac{\eE[\Delta\X(t)\,|\,\X(t)=\X]}{\Delta t},\quad
 D\mb{I}=\lim_{\Delta t \rightarrow 0} \frac{\eE[\Delta\X(t)^T\Delta\X(t)\,|\,\X(t)=\X]}{2\Delta t}.
\eeq

In practice, the expectations are estimated by finite sample averages and $\Delta t$ is the
time-resolution of the recording of the trajectories, as described in \cite{HozePNAS}.  Thus the estimates \cite{Masson} seem overly optimistic as far as the relation between the theoretical and the sampled trajectories is concerned.
By definition, the moment estimation in \eqref{eq:force} requires small fragments of trajectories
passing through each point of the membrane surface, which is precisely the massive data generated by
the sptPALM method \cite{Manley,Giannone} on biological samples. The empirical estimator defined by
\eqref{eq:force} has been used over the past 60 years in signal processing \cite{Schuss3} and
recently applied in several cell biology contexts \cite{Verstergaard}.  {The empirical estimator used in \cite{Hoze2012} is optimal as its variance is minimal and equals to the Cram\'er-Rao lower bound, as shown in section IIIE of \cite{Verstergaard}, using numerical simulations. This approach is clearly different from the Bayesian  method (see formula 6 in \cite{Masson} and previous publications).}

A positive point of the manuscript \cite{Masson} is the attempt to include in the analysis the influence of the Gaussian instrumental noise on the localization of moving receptors and thus on their trajectories. { However, the variance term in formula 6 of \cite{Masson} is inapplicable to sort out the biophysical parameters when there is a force or a drift component that can vary significantly in space. This is indeed the case at potential wells. The additional term, needed in formula 6 of \cite{Masson} appears in the direct estimation of the variance $E\left( (\X_{n+1}-\X_n)^2|\X_n=\X \right)$ and is proportional to the Laplacian of a potential well $\Delta V$, multiplied by the variance of the localization precision.} Finally, the motion blur due to the open camera shutter during acquisition can also be included in the analysis, as recently discussed in \cite{Verstergaard}. \\[1mm]
\textbf{The potential energy} \\[1mm]
The fluctuation-dissipation principle invoked in \cite{Masson} expresses the  relation between the local friction acting on the diffusing receptor and the diffusion coefficient. This relation is microscopic and is a part of the collision model of diffusion. The observed coarse-grained dynamics, which depends on obstacle density and may vary spatially, while friction remains constant, should not be expected to obey the fluctuation-dissipation principle \cite{hoze}. Formula  1 of \cite{Masson} postulates that the friction coefficient is an average with respect to the obstacle density, which leads to a coarse-graining approximation, incompatible with the local interpretation that the potential creates by local molecular interactions (which can extend only  to few nanometers). This interpretation is thus inconsistent with the hundreds-of-nanometers size of the effective potential well described in Fig. 3a of \cite{Masson}. As described above, the effective field $\mb{b}(\X)$  that accounts for local traps \cite{taflia} may have no potential at all. The relation between the local molecular energy and the size and depth of the potential well remains unclear. \\[1mm]
\textbf{Energy maps} \\[1mm]
The purpose of estimating the drift term is to clarify whether forces other than diffusion contribute to the dynamics. In this respect, although a clear large locally confining potential well was presented in \cite{Masson}, it is not clear what is the meaning of the energy map presented in Fig. 2c and f.  Potential interactions are local and the force is due to the gradient of the potential, thus a large region with  given constant energy indicates pure Brownian motion in this region, without interactions with any field. Is the energy actually calibrated or extracted from data?\\[1mm]
\textbf{Simulations} \\[1mm]
Another unclear point of the paper \cite{Masson} is the role, goal, and the construction of the simulations. The statement "Fokker-Planck equations can always be approximated by master equations" seems to indicate that the observed drift vector and diffusion tensor, which may be space and time-dependent, are known in the entire domain for all relevant times, which may be quite long. The Fokker-Planck equation (FPE) can be extracted from its global empirical solution, which is not contained in the empirical data. Thus the only way to extract a FPE from the data is to estimate the coefficients from the data, as described in \cite{HozePNAS}. Simulations of long trajectories of the assumed diffusion process can be run once the coefficients have been estimated in the entire domain. The approximations of the solution of the FPE proposed in \cite{Masson} is a path integral that requires entire trajectories, not only disjoint fragments. The statement "these simulated trajectories [...] characteristics match those of the experimental ones"  should be clarified, because theoretically, all diffusion processes may have the same trajectories, but with different probabilities. Thus simulated trajectories do not "match" empirical trajectories. What can be matched are various statistics of the trajectories, though it is not enough to match moments, because statistics, such as first passage times, which may be rare events, cannot be extracted from short simulations and certainly not from short fragments of trajectories. { The computation of the MFPT of a given diffusion process to a boundary requires either solving the Pontryagin-Andronov-Vitt boundary value problem \cite{Schuss2} or running Brownian simulations, once the drift and diffusion tensor have been estimated. But it cannot be computed from the solution of the Fokker-Planck equation in free space, because absorbing boundary conditions are necessary. A new approach based on stochastic simulations in the entire dendrite is developed in \cite{HozeBJ2014}.}

{Finally, estimating biophysical quantities, such as the first and second moments from the path-integral approach { used in \cite{Masson} may} introduce another computational bias: {using entire trajectories instead of two consecutive points from the same trajectory to compute these moments results in an estimation procedure that smoothes out the values of the diffusion coefficient.} The diffusion coefficients and energy extracted in \cite{Masson} are significantly lower and smoother than the ones presented in \cite{HozePNAS}. It is not clear whether this is a consequence of the method of \cite{Masson} or it captures the biophysical reality of inhibitory receptor trafficking.}

\end{document}